# AC Heat capacity and magnetocaloric effect measurements for pulsed magnetic fields

Yoshimitsu Kohama,[1,(a)] Christophe Marcenat,[2] Thierry Klein[3,4], and Marcelo Jaime[1]

[1]*MPA-CMMS, Los Alamos National Laboratory, Los Alamos, New Mexico 87545, USA*

[2]*CEA, Institut Nanosciences et Cryogénie, SPSMS-LATEQS, 17 rue des Martyrs, 38054 Grenoble, France*

[3]*Institut Néel, CNRS, BP 166, 38042 Grenoble, France*

[4] *Institut Universitaire de France and Universitè Joseph Fourier, BP 53, 38041 Grenoble, France*

(a) Electronic mail: ykohama@lanl.gov

A new calorimeter for measurements of the AC heat capacity and magnetocaloric effect of small samples in pulsed magnetic fields is discussed for the exploration of thermal and thermodynamic properties at temperatures down to 2 K. We tested the method up to $\mu_0 H$ = 50 Tesla, but it could be extended to higher fields. For these measurements we used carefully calibrated bare chip Cernox® and $RuO_2$ thermometers, and we present a comparison of their performance. The monotonic temperature and magnetic field dependences of the magneto resistance of $RuO_2$ allow us to carry on precise thermometry with a precision as good as ±1 mK at $T$ = 2 K. To test the performance of our calorimeter, AC heat capacity and magnetocaloric effect for the spin-dimer compound $Sr_3Cr_2O_8$ and the triangular lattice antiferromagnet $RbFe(MoO_4)_2$ are presented.

## I.    INTRODUCTION

Thermal properties such as specific heat and entropy are fundamental material properties that help us to understand the most relevant microscopic mechanisms. Among them, the quick and reliable determination of magnetic field–temperature ($H$,$T$) phase diagrams is highly valuable in condensed matter physics, since it helps to establish the basic ingredients for minimalistic modeling and prediction. Many thermodynamic studies including specific heat ($C_p$) and magnetocaloric effect (MCE) in very high magnetic fields have been reported over the last decade.[1-13] However, these reports have all been limited to the measurements carried out in DC magnets and one pulsed magnet that can produce a 100-500 ms magnetic field plateau.

The Pulsed Magnetic Field Facility of the National High Magnetic Field Laboratory at Los Alamos National Laboratory currently provides pulsed magnetic fields up to 85 T of different duration, ranging from 0.025 to 2 s, with and without flat tops. These magnets are world-class tools that so far have yet to be fully utilized for thermal studies of materials at low temperatures. Here we describe the development of a method for measuring heat capacity in pulsed magnetic fields using

AC calorimetry techniques. Since this method was introduced in 1968 by Sullivan and Seidel,[14] many studies in organic crystals,[15] superconductor,[16-18] liquid crystals,[19] fluids[20] and biological materials[21,22] have been carried out. Especially, the application of AC calorimetry techniques in measurements under pressure stimulated intense research activities.[23-25] This arises from the fact that AC calorimetry technique exploits effective noise rejection strategies and presents a very high sensitivity. Hence, the application of the AC calorimetry for measurements in the demanding electromagnetic environment of high/fast pulsed magnetic fields provides a new avenue for high quality thermal and thermodynamic research of materials.

In addition to AC heat capacity (AC-$C_p$) our calorimeter can also be used to measure the magnetocaloric effect (MCE) to map out the full ($H,T$) phase diagram in an efficient way, *i.e.* revealing phase boundaries before an extensive AC-$C_p(H,T)$ experiment is run, and to directly quantify entropy changes at phase boundaries. To assess the potential and performance of this method, new results are compared with MCE and $C_p$ data recently obtained up to $\mu_0H = 35$ T in a DC resistive magnet in the spin dimmer compound $Sr_3Cr_2O_8$ and the triangular lattice antiferromagnet $RbFe(MoO_4)_2$. To the best of our knowledge, this is the first report of AC-$C_p$ and MCE measurements in 250ms pulsed magnetic fields.

## II. PRINCIPLES UNDERLYING AC-$C_p$ and MCE MEASUREMENTS

Heat capacity is a fundamental thermodynamic quantity which is defined as the amount of energy or heat ($\delta q$) required to increase the temperature of a material by the amount ($\delta T$), $C = \dfrac{\delta q}{\delta T}$. Using the definition of entropy ($S$), $TdS = \delta q$, the specific heat can be expressed as, $C = T\left(\dfrac{\partial S}{\partial T}\right)_H$. Thus, the magnetic contribution to the heat capacity of materials measures the change of the spin entropy as a function of temperature. It is known that the spin entropy on magnetic materials can be altered with applied external magnetic fields. The change in the spin entropy as a function of magnetic fields leads to a measurable MCE, which is generally recognized as the change in temperature through the application or removal of an external magnetic field. The general formula for the entropy change in temperature and magnetic fields is

$$dq = TdS = T\left(\frac{\partial S}{\partial T}\right)_H dT + T\left(\frac{\partial S}{\partial H}\right)_T dH. \qquad \textcolor{red}{Eq.(1)}$$

This equation can be rearranged as

$$P = C\frac{dT}{dt} + T\left(\frac{\partial S}{\partial H}\right)_T \frac{dH}{dt}, \qquad \textcolor{red}{Eq.(2)}$$

where $P$ is power applied to the system $P = \dfrac{dq}{dt}$. The second term in the right hand side represents the power generated by the entropy change in field, which leads to a MCE.

The essence of MCE is depicted in Figure 1. Indeed, Figure 1(a) shows the entropy vs temperature for a magnetic system in zero magnetic field and in a finite applied magnetic field. The nature of the example material is arbitrary, with the only requirement being that the system entropy is reduced ($\frac{\partial S}{\partial H} < 0$) by the application of a magnetic field. When we increase the applied magnetic field in adiabatic conditions, as indicated by the dashed line in Fig 1(a), the sample temperature must increase from A to B. When the magnetic field is decreased back to zero in the same adiabatic conditions the sample temperature decreases from B to A. The previous statements are true disregarding of the presence or not of a field-induced phase transition. Note that in the cases where the entropy of the system increases with applied magnetic field, such as when charge, spin, or crystal electric field energy gaps are closed, the sign of the temperature change is simply inverted. Depending on the sweep rate of the magnetic field $dH/dt$ and the strength of the thermal link between sample and thermal bath during the MCE experiment, three clear situations are possible. Using the thermal conductance of the thermal link $\kappa$, $P$ can be expressed as the transferred heat from the thermal bath $P = \kappa\left(T - T_{bath}\right)$ and we can introduce a general formula for MCE measurements,

$$\kappa\left(T - T_{bath}\right) = \kappa\tau_1 \frac{dT}{dH}\frac{dH}{dt} + T\left(\frac{\partial S}{\partial H}\right)_T \frac{dH}{dt}, \qquad \text{Eq.(3)}$$

where $\tau_1$ is called the external time constant, $\tau_1 = C_p/\kappa$. If $dH/dt$ is extremely slow compared to $\tau_1$ and/or $\kappa$ is large, the experiment is performed in isothermal conditions ($T$-$T_{bath} \cong 0$), and no changes in the sample temperature vs magnetic field are observed as shown in Fig 1(b), where the red and blue lines represent data take during field up-sweep and field down-sweep. On the other hand, if the experiment is performed in adiabatic conditions ($dH/dt$ is quite fast and/or $\kappa$ is small), the sample temperature changes most rapidly, see Fig 1(c). In this case the sample temperature increases during the up-sweep, in agreement with Fig 1(a), and decreases during the down-sweep. The temperature change of the sample in this case is completely reversible. Figure 1(d) displays the most common practical case, where the experiment is performed in quasi-adiabatic conditions ($dH/dt$ and/or $\kappa$ are intermediate values). Here the sample over-heats during the field up-sweep, and then returns to the bath temperature within a time comparable to $\tau_1$. Since it over-cools during the field down-sweep, an open loop structure in MCE is observed. This last experiment is the most useful of all three limits to determine energy scale crossovers. In the typical MCE experiments in DC fields ($\tau_1 \sim$ 1s, $dH/dt \sim$ 0.01 T/s, $dT/dH \sim$ 0.1 K/T, $T$-$T_{bath} \sim$ 0.1 K),[4-7,15] the contribution of the first term in the right hand side of Eq.3 is much smaller than the term in the left hand side, hence it can be ignored. The resultant formula allows us to evaluate the change of the entropy as a function of magnetic field.[26]

The effects in the MCE are greatly enhanced when the applied magnetic field induces phase transitions, disregarding of the order of the phase transition. This phenomenon is particularly pronounced at temperatures and fields corresponding to magnetic phase transitions, schematically described in Fig. 1 (e)-(f), and thus it is a powerful tool for the investigation of the magnetic state and mechanisms of these magnetic transitions.[8,9, 11-13,26] The shape of the MCE response depends on the absence/existence of dissipative processes at crossover fields or phase boundaries (reversible phase transitions in Fig. 1 (c)-(e)). If an ideal reversible phase transition occurs, a symmetric temperature change in the up- and down-sweeps such as in Fig. 1 (d)-(e) is observed. However, if dissipative processes take place, such as domain wall motion, non-symmetric temperature changes can be observed manifesting itself as different amplitudes in the temperature changes during up-sweep and down-sweep as schematically displayed in Fig. 1 (f). The MCE data can then be used to separate reversible from irreversible processes at phase boundaries in magnetic materials, although making a clear distinction between transitions of different order is much more difficult.[9] Contrary to the case of properties measured as a function of temperature, hysteresis in MCE measured vs field is NOT (as clearly seen in the qualitative figures) a measurement of latent heat, and hence NOT a direct evidence for a 1st order phase transition. Sharpness of the MCE anomaly, in similar way, can be varied with d$H$/d$t$, $\tau_1$, $\kappa$ and so on, and NOT as often suggested indication of a 1st order phase transition. The only property that MCE reveals unambiguously is the presence or not of dissipation/irreversibility at the phase boundary, which can sometimes be linked to the order of the phase transition.[9]

In the AC-$C_p$ measurements, $P$ in Eq.2 contains the term of the applied AC power to the heater, $P = \kappa\left(T - T_{bath}\right) + P_{AC}\exp(i2\omega t)$, where the $P_{AC}$ is generated by the AC current of frequency $\omega$, and the sample temperature is $T = T_{bath} + T_{DC} + T_{AC}\exp(i2\omega t)$. Substituting $P$ and $T$ into Eq.2 gives the AC component

$$P_{AC} = \kappa T_{AC} + i2\omega C T_{AC} + T_{AC}\left(\frac{\partial S}{\partial H}\right)_T \frac{dH}{dt}, \qquad \text{Eq.(4)}$$

and then the time-dependent temperature can be write down as

$$T_{AC} = \frac{P_{AC}}{\kappa + \left(\frac{\partial S}{\partial H}\right)_T \frac{dH}{dt} + i2\omega C}. \qquad \text{Eq.(5)}$$

By solving this complex function, Eq.5 yields the real part of the denominator

$$\kappa + \left(\frac{\partial S}{\partial H}\right)_T \frac{dH}{dt} = \frac{P_{AC}}{\left|T_{AC}\right|}\cos(\varphi), \qquad \text{Eq.(6)}$$

and the imaginary part

$$-2\omega C = \frac{P_{AC}}{\left|T_{AC}\right|}\sin(\varphi), \qquad \text{Eq.(7)}$$

where the phase difference ($\varphi$) between the temperature response and the periodic heating power is

$$\tan(\varphi) = -2\omega C \Big/ \kappa + \left(\frac{\partial S}{\partial H}\right)_T \frac{dH}{dt}. \qquad \text{Eq.(8)}$$

The equations 7 and 8 for the AC heat capacity should be applied in the appropriate measurement frequency range relates to main two time constants. One time constant is determined by the thermal diffusivity ($D$) and the thickness ($d$) of the sample, thermometer and heater ensemble. This is often referred as the internal time constant ($\tau_2$), which is associated with the temperature homogeneity inside the sample, thermometer and heater ensemble. The other time constant called external time constant ($\tau_1$) corresponds to the time required for the system to reach equilibrium with the thermal bath. The proper measurement frequency ($2f$) for Eq.7 should be short compared to $1/\tau_2$ ($2f \ll 1/\tau_2$) and high compared to $1/\tau_1$ ($1/\tau_1 \ll f$).[27] In this frequency range ($1/\tau_1 \ll 2f \ll 1/\tau_2$), $\varphi$ is close to -90 degree, $|T_{AC}|$ is inversely proportional to $C_p$, and $|T_{AC}|$ times the frequency ($|T_{AC}| \cdot f$) becomes constant as a function of $2f$, where $f$ is defined as $f = \omega/2\pi$ (*i.e.* real part in Eq. 6 is negligible). Thus, an observation of the plateau in the $|T_{AC}| \cdot f$ vs $2f$ plot is a strong certificate of this frequency condition.

It is quite important to emphasize here that ANY and ALL pulsed field experiments are affected by the MCE phenomenon, *i.e.* reversibility (the absence of loops) in physical properties measured as a function of magnetic field *does not necessarily imply that the sample temperature is constant during the magnetic field pulse*. Our AC-$C_p$ experiment under pulsed magnetic fields is no exception, and the effects of MCE might appear as a change in $\varphi$ and induce a change in the evaluated $C_p$. Hence, we usually check for MCE effects in $C_p$ by changing the magnetic field sweep rate d$H$/d$T$.

## III. EXPERIMENTAL SETUP

A cross-sectional view of the present calorimeter is sketched in Fig. 2. The calorimeter cell consists of a small bare-chip resistive thermometer and a Si block acting as thermal bath. The small bare-chip resistive thermometer was electrically contacted with constantan wires of 25.4 μm diameter (0.001 in) using a silver epoxy (EPO-TEK® H20E). The face on which electrical contacts were made was directly glued to the Si thermal bath using an epoxy resin (Stycast® 2850). Extra wires were glued to the Si thermal bath to reduce vibration-related electromagnetic noise. In our estimation, the thickness of the epoxy layer is about 150 μm which acts as a small thermal insulation between the thermometer and Si thermal bath. We tune the thermal conductivity between sample and thermal bath with a variable amount of $^4$He (or $^3$He) exchange gas. The small amount gas contributes with a small heat capacity of the addenda, which we could neglect. Because the thermal conductivity of a gas is determined by the mean translational kinetic energy and the mean free path,[28] its thermal conductivity is independent of the magnetic field. As resistive thermometers, we tested both Cernox® and RuO$_2$ bare chips in a similar arrangement. The RuO$_2$ resistive thermometer is a commercially available thin-film chip resistor (State of the Art, Inc. S0302DPG221F-W). This thermometer was polished on one side to a final dimension of $900 \times 500 \times 150$ μm$^3$. Silver paint (SPI no. 5002) and GE-7031 varnish were used as a thermal contact between sample and thermometer. In the case of AC calorimetry, we used silver paint due to its extremely high thermal conductivity, while in the case of MCE measurement, both contact materials worked well.

In this work, we choose $Sr_3Cr_2O_8$, $RbFe(MoO_4)_2$ and Si single crystals as test samples. The $Sr_3Cr_2O_8$ single crystal samples were grown using an optical floating zone method at McMaster University,[29] the $RbFe(MoO_4)_2$ single crystals were grown by a flux method,[30] and the Si single crystals are commercial samples from Insaco, Inc. A ~100 Å nickel-chromium (NiCr) layer was deposited on one side of the sample, to use as heater. This layer resistance is typically ~10 kΩ at 4 K, and shows small temperature and magnetic field dependence compared to any thermometer. For example, at 4K, a 1 % change in the temperature leads to only 0.05 % change in the resistance of the NiCr layer, which is 20 times smaller than that of $RuO_2$ thermometer. In our measurement, the magnetoresistance of NiCr is only 0.1 % up to 45 T, which is 50 times smaller than our $RuO_2$ thermometer. The magnetoresistance and temperature dependence of the $RuO_2$ thermometer, as well as that of a bare chip Cernox® thermometer are shown in the next section.

Figure 3(a) and 3(b) show the block diagram for the data acquisition setup used in the MCE and AC-$C_p$ measurement systems, respectively. In the MCE measurement, we used a four contact AC technique with digital lock-in system, part of NHMFL's routine measurements of the electric resistance in pulsed magnetic field.[31] We also measured the magnetoresistance of thermometers with a four contact DC technique, and observed no difference besides the signal to noise ratio between AC technique and DC technique in the down sweep. A 100 Ω shunt resistor was used to determine the AC current through the $RuO_2$ thermometer. For driving constant AC current, we used a 1 kΩ resistor and a 1:100 transformer with an in-house AC voltage source. Since the effective impedance in this circuit is quite high (~10 MΩ), this part successfully generates a constant AC current during pulse.

In the AC-$C_p$ (Fig. 3(b)) measurement, we drive an AC current of frequency $f$ through the NiCr film heater using an in-house AC voltage source, a resistor and a transformer as shown in Fig 3(b), which induce temperature oscillations of frequency $2f$ in the sample. The resultant temperature oscillation is detected with a DC technique. To reduce noise, we choose a commercial battery equipped with a ~400 kΩ resistor as a current source for the DC detection. In a separate experiment we confirmed that the change in the current during pulse field is negligible small. On some experiments a Keithley 220 current source was also used. The amplitude and the phase of AC current flowing in the heater were measured by means of a 100 Ω resistance. The oscillating signal in the thermometer was split and amplified via two Stanford SR560 preamplifiers, where one preamplifier measures the entire signal without filtering the DC component and another preamplifier filtered the DC component. After amplification by factors of 10 - 100, the signals were stored in our data acquisition computer by using commercial National Instruments® and GaGe Applied

Technologies® digitizers. In both measurements, the digitizers collected the time dependence of the signal using a data acquisition rate of $0.1 \sim 20M$ samples per second.

While we have performed, and reported,[26] MCE data taken in a 250 ms and 2 s pulsed magnet, for space reasons we limit our discussion here to measurements in a 50 T capacitor-bank driven pulsed magnet with 250 ms pulse duration. During the magnetic field up-sweep, the average rate is 50T/25ms = 2000 T/s, while the magnetic field in the down sweep changes much more slowly. The down-sweep rates, while being a function of the time, are approximately 300 T/s and 150 T/s when peak fields are 40T and 15 T respectively.

## IV. MAGNETORESISTANCE of Cernox® AND RuO$_2$ THERMOMETERS

When performing AC-$C_p$ and MCE measurements in high magnetic fields at low temperature, it is quite important to correctly account for the magnetoresistance of the resistive thermometer. However, a number of reports on resistive thermometers have been limited to magnet field range generated by DC magnet.[32,33] In this section, we shortly report the magnetoresistance of RuO$_2$ and Cernox thermometers measured in a pulsed magnet, and describe an advantages of RuO$_2$ thermometer for AC-$C_p$ and MCE experiments.

Figure 4(a) and (b) show three-dimensional (3D) plots of magnetoresistance up to ~6 K for a Cernox® thermometer (CX-1010) and a RuO$_2$ thermometer. Both thermometers show clear magnetoresistance with decreasing temperature. The CX-1010 thermometer exhibits complex (mixed positive and negative) magnetoresistance below 10K. For example, at about 3 K, the magnetoresistance shows small bump at low fields. Similar magnetoresistance has been reported previously.[32,33] On the other hand, the RuO$_2$ thermometer shows monotonic positive magnetoresistance, and its relative change ($R$(45T)/$R$(0T)) only changes from 5% to 6% in the same measurement temperature range. Additionally, while the RuO$_2$ slightly gains temperature sensitivity (d$R$/d$T$ becomes larger) at high fields, the CX1010 looses significant d$R$/d$T$ at similar high fields. These features favor the RuO$_2$ resistance for utilization as a thermometer in high magnetic field research. On the other hand, while Cernox® thermometers behave nicely at high frequencies, RuO$_2$ thermometers show somewhat strong frequency dependence and thus we typically choose relatively low frequencies (1-2 kHz) for our measurements. The observed frequency dependence (1% at 5 kHz) might be an intrinsic feature of RuO$_2$ caused by the dielectric glass matrix in the microscopic structure.[34] When using frequencies above ~ 3 kHz, it is then necessary to re-calibrate the magnetoresistance of RuO$_2$ at each frequency.

The relatively small and monotonic magnetoresistance in the RuO$_2$ thermometer allows us to fit the magnetoresistance with the simple polynomial function:

$$R_{fit}(H,T) = a_0 + a_1 H^{0.5} + a_2 T^{-1} + a_3 H + a_4 T^{-2} + a_5 H^{0.5} T^{-1} + a_6 H^{1.5} + a_7 T^{-3} + a_8 H^{0.5} T^{-2} + a_9 H T^{-1},$$

where $a_n$ for $n = 0, 1, \ldots 9$ are the polynomial coefficients. As shown in Fig. 4(b) with a red surface, the fitting function has an excellent agreement with data. The deviation of the fitting function from the data points is within ~0.2 % in the entire temperature and magnetic field ranges. This error corresponds to ~20 mK, ~10 mK and ~4 mK at 6 K, 4 K and 2 K, respectively.

## V. MAGNETOCALORIC EFFECT RESULTS

To test advantages and shortcomings of MCE and AC-$C_p$ measurements in pulsed magnet field, we used a single crystal sample of the spin dimmer compound $Sr_3Cr_2O_8$ which shows XY-type antiferromagnetism between $H_{c1} = 30.4$ T and $H_{c2} = 62$ T. The ($H_c$, $T_c$) phase boundary for this quantum paramagnet system has a dome shape that is limited to $T_{cmax} \sim 8$ K in temperature,[26] qualitatively similar to other systems previously studied.[5,11,13] This family of compounds has a non magnetic spin-singlet ground state, with a finite energy gap to the first excited spin-triplet states. A strong enough external magnetic field can be used to close the spin gap, inducing magnetic ordering. Hence, because of the closing of the gap and the induced magnetic ordering, $\Delta S$ shows a peak structure as a function of magnetic field which is discussed in the work by Aczel et al.[26] The MCE results expected in this system include a crossover-type behavior discussed in Fig 1(d), with the opposite sign because $S$ increases with field in our sample, and also the phase transition-type of Fig. 1(e). This behavior has been observed before in $Ba_3Mn_2O_8$ at lower fields produced with DC resistive magnets.[11]

We show MCE data for $Sr_3Cr_2O_8$ taken at $T = 2.4$ K and 4.5 K in Fig. 5 (a). Here, the sample dimensions are $300 \times 350 \times 100$ μm$^3$. The up-sweep data show broader downward peak, while the down-sweep data show upward sharp bump. At lower temperature the MCE responses become small due to the smaller entropy change at the phase boundary. In our data, the MCE curve clearly consists of two typical MCE responses described in Fig. 1(d) and 1(e) superimposed. Below ~30 T, the MCE response corresponds to the entropy increase due to the closing of the spin gap. Above ~30 T, a careful look of the up-sweep and down-sweep data taken at 4.5 K reveals a sharper temperature change which is related to the 2$^{nd}$ order magnetic phase transition. Although a quantitative evaluation of $S$ is difficult, the symmetric response indicates that a reversible process is predominant in this magnetic phase transition. In fact, the MCE measurement in DC field shows clear symmetric MCE at $H_{c1}$.[26] The evaluation of $H_{c1}$ from pulsed field data can be done by taking the first derivative of temperature vs field $\partial T/\partial H$ as shown in Fig. 5(b). One peak is observed at each sweep, and the peak position in the up-sweep is higher than that in the down-sweep. The difference between up and down-sweeps can be explained by the sample lagging behind during the fast up-sweep, which also observed in the experiments using DC magnet with high sweep rates. This is because the sample temperature needs a finite time characterized by $\tau_1$ to respond to the heat released from the entropy change, like in traditional thermal relaxation time type calorimetry. Since the delay should be proportional to the sweep rate ($dH/dt$), the critical magnetic field $H_c$ can be evaluated from the following equation:

$$H_c = \left( H_{c,peak-down} \frac{dH}{dt}_{up} + H_{c,peak-up} \frac{dH}{dt}_{down} \right) \bigg/ \left( \frac{dH}{dt}_{up} + \frac{dH}{dt}_{down} \right).$$    (Eq.10)

As a result, $H_{c1}(T)$ is evaluated as 30.8 T at 2.4 K and 32.1 T at 4.5K, where the departure from the evaluated phase diagram using DC magnet is smaller than 1 T.[26] We believe that the main source of error is the delay in the sample response ($\tau_1$ not small enough) and the small number of data points. The use of a smaller sample and a measurement at higher frequency can reduce the error. As an example, the derivative data at 4.2 K measured at 10 kHz with smaller sample size ($200 \times 300 \times 50$ μm$^3$) are plotted in the inset of Fig. 5 (b). The clean data obtained this way show a smaller separation between the up-sweep and down-sweep peaks, which allows us to obtain the phase boundary more precisely.

## VI. FREQUENCY TEST AND NUMERICAL SIMULATIONS

In MCE measurements, the delay of the thermal response leads to broad MCE curves and makes difficult to evaluate the correct field-temperature phase diagram in magnetic materials. The time scale of the delay could be the "external" time constant ($\tau_1$) as described in the previous chapter, which is typically a few orders of magnitude larger than the "internal" time constant ($\tau_2$). If we apply the AC calorimetric technique, we can dramatically reduce the time scale of the delay from $\tau_1$ to $\tau_2$ and could earn more precise phase diagram. In addition, the evaluation of $C_p(H)$ gives us a chance to discuss the implications of the physical pictures.

In Fig. 6, we plot experimental results of $|T_{AC}| \cdot f$ vs $2f$ and $\varphi$ for Sr$_3$Cr$_2$O$_8$ and Si single crystals, which were measured at ~4 K and zero magnetic field. In this frequency test, the sample dimensions for the Sr$_3$Cr$_2$O$_8$ and Si single crystals are $450 \times 250 \times 100$ μm$^3$ and $450 \times 200 \times 130$ μm$^3$, which have heat capacity of $1.1 \times 10^{-8}$ and $5.4 \times 10^{-10}$ JK$^{-1}$, respectively. The Sr$_3$Cr$_2$O$_8$ sample shows a clear plateau in $|T_{AC}| \cdot f$ between 1 to 10 kHz, while the Si sample only exhibits two kinks due to its small contribution to the total heat capacity. The phase $\varphi$ in both samples are close to -90 degree near 2 kHz. In the Sr$_3$Cr$_2$O$_8$ sample, the frequency dependences observed in $|T_{AC}| \cdot f$ and $\varphi$ indicate that the frequency range near 2 kHz (for $2f$) closely satisfy the requirements ($1/\tau_1 \ll 2f \ll 1/\tau_2$) for the AC-$C_p$ measurement. In the case of the Si sample, we should keep in mind that the addenda contribution to the total heat capacity can decrease with increasing measurement frequency, even if the frequency condition ($1/\tau_1 \ll 2f \ll 1/\tau_2$) is fulfilled, because the variation of the thermal length with changing measurement frequency can affect the addenda heat capacity: This addenda heat capacity is originated in the He exchange gas, heater, thermometer, contact, electrical wires and Stycast. Therefore, the small heat capacity of Si comparing to the addenda heat capacity might explain no plateau behavior in the Si frequency test result.

In order to understand the specific features of our AC calorimeter and the frequency test results in Si, we solved a simple 1D heat equation model using a finite element method, which is commonly used for finding approximate solutions for the heat equation.[35] The homemade computation program was built in a LabVIEW™ 8.2 package from National Instruments. The vertical arrows on the right hand side of Fig. 2 represent the construction of our model, where the model consists of 6 regions, [4]He gas, sample, contact, thermometer, thermal insulation layer, and Si thermal bath. For simplicity, we neglected the thin RuO$_2$ layer, NiCr layer, and the constantan wires. We chose 30 μm thickness and 0.1 WK$^{-1}$m$^{-1}$ thermal conductivity for the contact layer (silver paint).[36] Other physical values used in this calculation, such as heat capacity and thermal conductivity in the each region, are taken from the literature,[36-48] which are listed in the figure caption of Fig. 7. Figure 7(a) shows the simulated temperature profile within these 6 regions at $2f$ = 1 kHz. The thermal oscillations caused by the AC power transfers from the top of the sample to the end of thermometer equipped RuO$_2$ thin layer, although the contact layer reduces by 10 % the thermal oscillation amplitude at this frequency. In the region of the thermal insulation layer (Stycast), the temperature oscillation is dumped and then the high thermal conductance Si thermal bath does not show any thermal oscillations. Since we observe the heat capacity in the region where the temperature is oscillating, this result confirms that our calorimeter measures the heat capacity of the sample, contact layer, thermometer, and a layer of of [4]He gas and thermal insulation layer. The heat capacity in [4]He gas (173.9 JK$^{-1}$m$^{-3}$ at 2.5 mbar),[48] silver paint (2400 JK$^{-1}$m$^{-3}$),[45] and Stycast (1400 J$^{-1}$K$^{-1}$m$^{-3}$)[44] cannot be neglected, when we measure a small specific heat sample such as Sr$_3$Cr$_2$O$_8$ (1000 J$^{-1}$K$^{-1}$m$^{-3}$)[26] and especially Si (45.7 J$^{-1}$K$^{-1}$m$^{-3}$).[46] However, since we can write down the measured heat capacity ($C_p^{total}$) as $C_p^{total} = C_p^{addenda} + C_p^{Sr_3Cr_2O_8}$ or $C_p^{addenda} + C_p^{Si}$, the addenda heat capacity can be eliminated by the following equation

$$\frac{P_{AC}^{Sr_3Cr_2O_8}}{2\omega \cdot T_{AC}^{Sr_3Cr_2O_8}} - \frac{P_{AC}^{Si}}{2\omega \cdot T_{AC}^{Si}} \approx C_p^{Sr_3Cr_2O_8} - C_p^{Si} . \qquad \text{(Eq.11)}$$

The variables on the left hand side are observables and we can determine $C_p^{Sr_3Cr_2O_8} - C_p^{Si}$. Since $C_p^{Si}$ is negligible small and also already reported elsewhere,[46] we successfully evaluated the $C_p$ of the sample in pulsed magnetic fields.

Figure 7 (b) and (c) are the calculated $|T_{AC}|$:$f$ vs $2f$ curves in Sr$_3$Cr$_2$O$_8$ and Si samples. The shapes of the curves in $|T_{AC}|$:$f$ and $\varphi$ are similar to the experimental result in Fig. 6(a) and (b). The simulated $|T_{AC}|$:$f$ curve in Sr$_3$Cr$_2$O$_8$ sample shows a slightly rounded plateau between 1 and 10 kHz. The clear shape of the plateau in the experiment seems to stem from the contribution of the extra silver paint to the total heat capacity, because the larger heat capacity can increase $\tau_1$ and extends the size of the plateau. The phase difference $\varphi$ in $Sr_3Cr_2O_8$ is close to -90 degree near 2 kHz, as like as the simulation result in Fig. 6, which is the desired condition in AC calorimetry. On the other hand, the $|T_{AC}|$:$f$ in the Si sample shows only kinks at 500 Hz and 10 kHz and $\varphi$ is close to -90 degree near 4 kHz. The absence of plateau could be caused by the change of the addenda heat capacity with frequency, because the contribution of the Si to total heat capacity is much smaller than the heat capacity from the addenda. As shown in Fig.7 (b), the $C_p$ change in the sample can be detected as a decrease of the $|T_{AC}|$, but the thermal conductivity change from 1 to 25 W/K$^{-1}$m$^{-1}$ does not affect $|T_{AC}|$. This suggests that our setup can measure sample heat capacity, but not its thermal conductivity.

Moreover, we can conclude that the damping in $|T_{AC}|$ happens in a low thermal diffusivity layer that is likely the thermal contact layer in our calorimeter. Even if some temperature damping exists, we still can measure $C_p$ by evaluating how the $|T_{AC}|$ is reduced in the contact layer. In the low frequency limit ($f \ll \kappa/4\pi C_p L^2$) where $L$ is the thickness of a layer, M. Ivanda et al., give an expression of the temperature oscillations of a low thermal diffusivity layer sandwiched by the heater and another low thermal diffusivity layer.[49]

$$T(x) \approx (A_1 + A_2 \exp(i\omega t)) \frac{\kappa_T |x| + \kappa_C L_T}{\kappa_T L_C + \kappa_C L_T}, \tag{Eq.12}$$

where $A_1$ and $A_2$ are constants $\kappa_T$, $L_T$ and $\kappa_C$, $L_C$ are the thermal conductivity and the thickness of each low thermal diffusivity layers. By neglecting the temperature gradients in the sample and thermometer layers, which commonly have higher thermal diffusivities than that of the contact layer, the equation approximates the temperature oscillation of the thermal contact layer in our AC calorimeter. In this case, $\kappa_T$, $L_T$ and $\kappa_C$, $L_C$ corresponds to the thermal conductivity and the thickness of the thermal insulation and contact layers in our calorimeter, respectively. By taking the ratio of two difference positions $L_C$ and 0 on the contact layer, we can get

$$\frac{|T_{AC}(0)|}{|T_{AC}(L_C)|} \approx \frac{\kappa_C L_T}{\kappa_T L_C + \kappa_C L_T}. \tag{Eq.13}$$

Then, Eq. 7 can be rewritten as $|T_{AC}| \sim -aP_{AC}/2\omega C \sin\varphi$ where $a$ equal to the right hand side of Eq. 13. Since $\kappa_T L_C$ in Eq. 13 is more than ten times smaller than $\kappa_C L_T$ in our AC-calorimeter ($\kappa_T \sim 0.04$ WK$^{-1}$m$^{-1}$,[38] $L_T \sim 150$ μm, $\kappa_C \geq 0.1$ WK$^{-1}$m$^{-1}$,[36] and $L_C \sim 30$ μm,), $a$ is close to 1 and expected not to vary dramatically with changing $\kappa_T$ and $\kappa_C$ in magnetic fields. For example, if we assume a 50 % decrease in $\kappa_C$ by applying external magnetic fields, likely an overestimate for the polycrystalline silver paint, the expected change in $a$ is only 7%. Thus, we can neglect the change in $a$ during pulsed magnetic fields.

## VII. AC-$C_p$ RESULTS

Fig. 8(a) shows the overall temperature change during pulsed magnetic field recorded in the AC-$C_p$ measurement with a Si single crystal. The red and black curves are data taken in the up and down-sweeps, respectively, and the blue curve is smoothed data. The $2f$ temperature oscillations (1 kHz) are clearly observed in both the up and down-sweep data (see also in the enlarged inset figure). Because of the lower $dH/dt$, the down-sweep curve shows a large number of temperature oscillations than the up-sweep curve. The heating effect due to the high $dH/dt$ in the up-sweep results in a small temperature increase at about 5 T. Except the low field region of the up-sweep, almost no heating effect is detected, which indicates that MCE is negligible small.

By multiplying the sinusoidal reference functions of frequency $2f$ by the observed signal, as is usually done in a commercial digital lock-in amplifier, we numerically extracted the $|T_{AC}|$ component after integration over one period without any further smoothing process. Figure 8(b) provides the measured $\dfrac{P_{AC}}{2\omega \cdot |T_{AC}|}$ in the Si sample at ~4.2 K in a pulsed field up to 43 T, which correspond to

$C_p^{addenda} + C_p^{Si}$. The black curve is evaluated from the raw temperature data of the down-sweep magnetic fields, and the blue curve is its smoothed curve. In these curves, one broad bump at low magnetic fields can be observed. The size of the bump is only ~10 nJK$^{-1}$ at ~5 T and its broad shape reminds us of a Schottky anomaly. In fact, the measurement temperature of ~ 4 K corresponds to the energy gap of free spins caused by Zeeman splitting.[50] Therefore, a tiny amount of free impurity spins contained in Si and/or calorimeter seems to be responsible for this broad anomaly. Above 20 T, the data is almost independent of magnetic fields. In this field region, the magnitude of heat capacity is two orders of magnitude higher than the expected heat capacity of Si (5.4×10$^{-10}$ JK$^{-1}$). This confirms that the addenda heat capacity is the major contribution in this measurement. Indeed, specific heat of Stycast [44] and Silver paint [45] are two orders of magnitude bigger than that of Si.[46] Hence, we can use the measured $\dfrac{P_{AC}}{2\omega \cdot |T_{AC}|}$ in this Si measurement as a heat capacity of addenda and can evaluate the sample heat capacity with Eq. 11. Fig. 8 (c) displays the deviation of the data points from the smoothed curve as a function of the magnetic fields. With increasing the magnetic fields, the dispersion of data becomes slightly larger due to the electronic noise, but it is still within ± ~0.5 %. At higher frequency, the dispersion is increased due to the smaller $|T_{AC}|$. For example, without smoothing process, the dispersions at 43 T are ± 1 %, ± 5 % and ± 15 % at $2f$ = 2, 5, and 10 kHz, respectively. With a large number of data points in the high frequency region, smoothing can effectively reduce the dispersion of the data points.

In order to confirm that our calorimeter has the sensitivity to detect heat capacity anomalies as a function of temperature and magnetic fields, we measured the triangular lattice antiferromagnet system RbFe(MoO$_4$)$_2$. In zero field, this sample shows a quite large heat capacity anomaly (3.8 ×10$^5$

JK$^{-1}$m$^{-3}$ at ~3.74 K [51] and 5.7 ×10$^5$ JK$^{-1}$m$^{-3}$ at 3.90 K [52]), and by applying external magnetic fields, several magnetic ordered states are induced.[51-53] Since the detected signal $|T_{AC}|$ is inversely proportional to the heat capacity, the AC-$C_p$ measurement of RbFe(MoO$_4$)$_2$ is quite difficult. The sizes of samples mounted on our calorimeter are 130 × 250 ×50 μm$^3$ (sample 1) and 120 × 330 ×70 μm$^3$ (sample 2) which have much larger heat capacity (6.2-15.9 × 10$^{-7}$ JK$^{-1}$ at peak temperature) than the addenda heat capacity shown in Fig. 8.

The black and red curves in the inset of Fig. 9(a) are $|T_{AC}|$:$f$ vs 2$f$ in the sample 1 and 2, respectively. Both samples show a peak at around 100 Hz which are quite different from those in Sr$_3$Cr$_2$O$_8$ and Si single crystals. The origin of the difference can be explained by the large specific heat of RbFe(MoO$_4$)$_2$ resulting in large $\tau_1$ and $\tau_2$, These large time constants shift the plateau to low frequencies and are able to change the shape of plateau. Although the absence of a plateau makes it difficult to evaluate the absolute value of the heat capacity, we tested the AC-$C_p$ measurements at 2$f$ = 400 Hz which is the lowest frequency limit of our setup. Figure 9(a) depicts the temperature dependence of $\dfrac{P_{AC}^{RbFe(MoO_4)_2}}{2\omega \cdot \left| T_{AC}^{RbFe(MoO_4)_2} \right|} - \dfrac{P_{AC}^{Si}}{2\omega \cdot \left| T_{AC}^{Si} \right|}$, which was taken with the temperature ramp rate of 2.5 Ks$^{-1}$ at 4 K. The black, blue and red curves were measured with the different AC power of 5.8, 8.3 and 16 μW, respectively. The data taken by the traditional relaxation method [52] (green circles) are in the right axis. Strictly speaking, $\dfrac{P_{AC}^{RbFe(MoO_4)_2}}{2\omega \cdot \left| T_{AC}^{RbFe(MoO_4)_2} \right|} - \dfrac{P_{AC}^{Si}}{2\omega \cdot \left| T_{AC}^{Si} \right|}$ in the left axis does not corresponds to $C_p^{RbFe(MoO_4)_2} - C_p^{Si}$, because no plateau behavior in $|T_{AC}|$:$f$ vs 2$f$ makes difficult to apply Eqs. 7 and 11 in this measurement. In fact, the evaluated heat capacity of RbFe(MoO$_4$)$_2$ becomes 6-9 times larger than the heat capacity values measured using the traditional thermal relaxation time technique. This is because the absence of a plateau reduces $\left| T_{AC}^{RbFe(MoO_4)_2} \right|$, and the $\dfrac{P_{AC}^{RbFe(MoO_4)_2}}{2\omega \cdot \left| T_{AC}^{RbFe(MoO_4)_2} \right|}$ term becomes much larger than the true heat capacity. However, $\dfrac{P_{AC}^{RbFe(MoO_4)_2}}{2\omega \cdot \left| T_{AC}^{RbFe(MoO_4)_2} \right|}$ still could follow the relative changes of the heat capacity. Indeed, the three curves measured by our AC calorimeter agree with the $C_p$ data in the Ref. 52. The agreement between data sets tells us that the main advantage of our calorimeter is its high measurement frequency and sensitivity. These allow us to complete quick measurements during fast temperature and magnetic field sweeps. In fact, each curve in Fig. 9(a) was taken within ~1 second. However, a limitation of AC-power is expected in the case of samples showing very large heat capacity and/or bad thermal conductivity due to the $\tau_2$ problem. Actually, the peak temperatures in the three curves (3.81 K, 3.92 K, and 3.94 K at 16, 8.3 and 5.8 μW, respectively) are shifted to lower temperature when a higher AC power is applied to the heater. This is due to the existence of a temperature gradient between the thermometer and the sample. This temperature gradient can also be found in the numerical results in fig. 7(a), where the application of $5 \times 10^{-10}$ W per μm² grid (corresponding to 16.25 μW in sample 1) gives an average temperature difference of ~0.15 K at the silver paint contact layer. Thus, a small AC power must be applied to avoid large temperature gradients. In the following measurements of RbFe(MoO₄)₂ in pulsed fields, we choose 8.3 μW for $P_{AC}$.

Figure 9(b) shows $\dfrac{P_{AC}^{RbFe(MoO_4)_2}}{2\omega \cdot \left| T_{AC}^{RbFe(MoO_4)_2} \right|} - \dfrac{P_{AC}^{Si}}{2\omega \cdot \left| T_{AC}^{Si} \right|}$ vs magnetic field up to 20 T. Here, the black,

blue and red curves are measured for different maximum magnetic fields of 20, 15, 10 T, which have different sweeping rates of magnetic field d$H$/d$t$ of 47, 35 and 23 T/sec at 5 T, respectively. Since

$\dfrac{P_{AC}^{RbFe(MoO_4)_2}}{2\omega \cdot \left| T_{AC}^{RbFe(MoO_4)_2} \right|} - \dfrac{P_{AC}^{Si}}{2\omega \cdot \left| T_{AC}^{Si} \right|}$ seem to be roughly proportional to the heat capacity in Fig. 9(a), we

labeled it as $C_p^{RbFe(MoO_4)_2}$ with arbitrary units. In the whole measurement temperature and magnetic fields range, $C_p^{RbFe(MoO_4)_2}$ shows pronounced anomalies. All of the observed anomalies in this measurement reproduce nicely previous heat capacity data taken using the traditional relaxation method,[53] and follow the known phase diagram.[51-53] Here, we want to point out that the three curves taken with changing d$H$/d$t$ show the same shape of the peaks especially at the high sensitive low field region. This means the change of $\left( \dfrac{\partial S}{\partial H} \right)_T \dfrac{dH}{dt}$ in Eq. 4 and 8 is negligible small and does not affect the Eq. 7 in these pulsed field measurements. The low sensitivity at high magnetic fields especially near peak positions is caused by the electric noise in pulsed fields, which is not negligible comparing to the detected small $|T_{AC}|$ of ~2 mK in this measurement. However, typical samples have much smaller heat capacity than RbFe(MoO₄)₂ and should show larger $|T_{AC}|$ and higher sensitivity. The AC-$C_p$ data in RbFe(MoO₄)₂ with temperature and magnetic field sweeps indicate that our calorimeter efficiently maps out the field-temperature phase diagram for magnetic materials.

Fig. 10(a) and (b) show the temperature oscillations at $2f = 1$ kHz during an AC-$C_p$ measurement with a Sr₃Cr₂O₈ single crystal. The red and black curves taken during up and down sweeps show the AC temperature oscillations. The amplified signal in Fig. 10(b) and its inset clearly detect the AC temperature oscillations with a high sensitivity of ± 2 mK at ~30 T. The averaged raw signal (blue curve) in Fig. 10 (a) exhibits the MCE at 33.1 and 32.1 T in the up and down-sweep. These magnetic fields are consistent with MCE data in pulsed magnetic field (see in Fig. 5), and the evaluated transition magnetic field with Eq. 10 agree with the previous $C_p$ and MCE data in DC field.[26] The agreement observed in the determination of critical magnetic field indicates that the temperature gradient between sample and thermometer caused by the AC power is negligible small in this measurement. By numerically extracting $|T_{AC}|$ from the temperature oscillation data in Fig. 10(b), we evaluated $\dfrac{P_{AC}^{Sr_3Cr_2O_8}}{2\omega \cdot \left| T_{AC}^{Sr_3Cr_2O_8} \right|} - \dfrac{P_{AC}^{Si}}{2\omega \cdot \left| T_{AC}^{Si} \right|}$, which is simply refereed as $C_p^{Sr_3Cr_2O_8}$ in Fig. 10(c). This data show a pronounced peak and a small dip at 32.7 and 32.1 T, respectively. The peak shape agrees with the $C_p$ data measured in DC magnetic fields (green circles)[26] which again confirms the reliability of our AC calorimeter. The dips indicated by the arrows, however, seem to be artificial signals caused by the MCE. One possible explanation of dips is the $\left( \dfrac{\partial S}{\partial H} \right)_T \dfrac{dH}{dt}$ term in Eq.8, which could affect $|T_{AC}|$ and $C_p$. Another explanation is the errors in the numerical evaluation process of $|T_{AC}|$ which might occur when the raw temperature signal shows a sudden jump. In fact, the size of jumps seen in Fig. 10 (b) and its inset are compatible with the $|T_{AC}|$ which might bring an artificial noise. In any case, the size of dip is negligible when compared with the big peak at the spin-ordering phase transition. Figure 10(d) displays the $C_p^{Sr_3Cr_2O_8}$ measured from 2.3 to 5.6 K. The peaks at spin-ordering phase transition are clearly observed in all temperature regions and its temperature dependence again agree with the previous $C_p$ and MCE measurement.[26] The bigger dip evident at lower temperatures could be caused by a bigger MCE reported in experiment performed in DC fields[26] and the broad peak at 5 T should correspond to a Schottky anomaly caused by probable impurity spins in the Sr$_3$Cr$_2$O$_8$ sample. The consistencies of the peak magnetic fields with the previous report are satisfactory and demonstrate that our calorimeter works reliably in a 250 ms pulsed magnetic field.

Finally, we would like to mention a few issues and future plans to improve our calorimeter. The first issue of our AC-$C_p$ measurement is the evaluation of the contribution of the silver paint to the total heat capacity. This is originated from the difficulty to determine the amounts of silver paint used for the thermal contact to the thermometer and the electrical contact to the NiCr heater, which represents a non-negligible contribution to the total heat capacity. A second issue is that a deposited NiCr heater does not work on metallic samples due to the high conductivity of the sample itself. To solve these problems, the application of the optical heating technique should be most effective and it could increase the sensitivity of our calorimeter. These tests are currently under way.

ACKNOWLEDEMENTS

YK and MJ would like to thank D. F. Weickert and Y. Suzuki for early discussions of the AC-$C_p$ method, F. F. Balakirev for the help with digital lock-in measurements, and J. B. Betts and A. Migliori for technical supports during experiment. We are indebted to A. A. Aczel, G. Luke, A. Ya. Shapiro and L. A. Demianets for providing the $Sr_3Cr_2O_8$ and $RbFe(MoO_4)_2$ single crystals used for these proof-of-principle experiments. This work was supported by the National Science Foundation, the U.S. Department of Energy and the State of Florida.

FIGURE CAPTIONS

Fig. 1. (a) Schematic entropy (*S*) vs temperature (*T*) in a hypothetical system where the entropy is reduced by an applied magnetic field as it is observed in a paramagnet. (b) Temperature of our hypothetical sample when the magnetic field is changed in an isothermal fashion, allowing the full exchange of heat with the bath. (c) Same field sweep done in an adiabatic fashion. As the sample cannot exchange heat with the bath the temperature changes reversibly as indicated in (a) when the sample travel from A to B and back to A. (d) The most realistic case of quasi-adiabatic field sweep reveals changes in the sample temperature that permit the determination of the relevant field/energy scale of the sample under study. (e) When the quasi adiabatic field sweep is performed in the presence of a field-induced second order phase transition at the critical field $\mu_0 H_c$ the magnetocaloric effect reveals sharp features. (f) First order phase transitions where dissipative mechanisms are present show characteristic asymmetry due to release of heat in both directions of the magnetic field sweep.

Fig. 2. Schematic drawing of our AC calorimeter. The arrows in the right hand side represent the 6 different regions used in our 1D simple model for simulating the temperature profile in our calorimeter. (See Fig.7)

Fig. 3. Circuit diagrams for (a) MCE and (b) AC-$C_p$ measurements. (a) The ports labeled $V_{AC}$ and $P_{AC}$ detect the resistance and the AC current changes in the thermometer, respectively. The preamplifiers settle in the $V_{AC}$ port removes the noise with the band-pass filter. The band-pass filter blocks the signal with the frequency 100 times higher and smaller than the AC current frequency (*f*). (b) $V_{DC}$, $V_{AC}$ and $P_{AC}$ are the ports to detect the resistance change of thermometer, the amplified AC resistance oscillation of the thermometer, and the applied AC power to the NiCr heater during experiments, respectively. The band-pass filter in $V_{AC}$ port rejects the signal with frequencies far from the AC temperature oscillation frequency (2*f*).

Fig. 4. 3D plot of the magnetoresistance of the (a) CX-1010 and (b) RuO$_2$ thermometers at low temperatures in fields up to 50 T. The resistance of CX-1010 and RuO$_2$ were measured at 60 kHz and 2 kHz, respectively. The red surface plot is the result of the surface fitting using the polynomial function discussed in the text.

Fig. 5. (a) MCE data in Sr$_3$Cr$_2$O$_8$. The solid symbols are data taken during up-sweep. The open circles are the data of the down-sweep. (b) Numerical derivative of the MCE data. The peak in the up-sweep at 2.4 K data is not obvious due to the slow varying MCE response. The blue data in the inset were taken at 4.2 K using measurement frequency of 10 kHz, while the data in the main panel

were taken at 1 kHz. The arrows indicate the peak positions.

Fig. 6. Frequency dependence of $|T_{AC}|\cdot2f$ and $\varphi$ in $Sr_3Cr_2O_8$ (a) and Si (b) single crystals. The maximum value of $|T_{AC}|\cdot2f$ at the top of the dome is normalized to 1.

Fig. 7. (a) Simulated temperature profile in our calorimeter at different time. The 12 curves have $6/\pi$ phase interval of the AC heating power. In this calculation, we used specific heat of 45.7 (Si), 2400 (Silver paint), 8.63 ($\alpha$ $Al_2O_3$), 1400 (Stycast), 1000-4000 ($Sr_3Cr_2O_8$), 173.9 $JK^{-1}m^{-3}$ ($^4He$ at 2.5 mbar), and thermal conductivity of 100 (Si), 0.1 (Silver paint), 0.04 ($\alpha$ $Al_2O_3$), 0.04 (Stycast), 1-25($Sr_3Cr_2O_8$), 0.00767 $WK^{-1}m^{-1}$ ($^4He$). The applied AC power ($P_{AC}$) is $5 \times 10^{-10}$ W to 1 $\mu m^2$ area and the thermal conductance from the edge of the silicon platform is $5 \times 10^{-6}$ $WK^{-1}$. The time step and distance step are 50-5000 ns and 1 $\mu m$. (b) Simulated frequency dependence of $|T_{AC}|\cdot f$ and $\varphi$ in $Sr_3Cr_2O_8$ single crystal. The black dots, circles and squares are the $|T_{AC}|\cdot f$ with $C_p = 1000$ and $\kappa=5$, $C_p = 2000$ and $\kappa=5$, and $C_p = 4000$ and $\kappa=5$. The black dashed and dot curves are the result $C_p = 1000$ and $\kappa=1$ and $C_p = 1000$ and $\kappa=25$, respectively. The red dots, circles and squares are $\varphi$ with $C_p = 1000$ and $\kappa=5$, $C_p = 2000$ and $\kappa=5$, and $C_p = 4000$ and $\kappa=5$. (c) Simulated frequency dependence of $|T_{AC}|\cdot f$ and $\varphi$ in Si single crystal. The black and red dots are $|T_{AC}|\cdot f$ and $\varphi$, respectively.

Fig. 8. (a) Temperature vs magnetic field for the Si single crystal during the 250 ms pulse. The frequency of the temperature oscillation ($2f$) is 1 kHz. The signal was detected without filtering the raw signal. The black, red, and blue curves are the data of down-sweep, up-sweep, and its smoothed data. (b) Magnetic field dependence of $\dfrac{P_{AC}^{Si}}{2\omega\cdot\left|T_{AC}^{Si}\right|}$. The black, curve is extracted from the data of down-sweep without any smoothing process, and the blue curve is its smoothed curve. For the clarity purpose, we do not show the $\dfrac{P_{AC}^{Si}}{2\omega\cdot\left|T_{AC}^{Si}\right|}$ data extracted from the data of up-sweep, which does not have enough measurement frequency under the fast up-sweep magnetic field and show a noise at about 5 T due to the small temperature increase as shown in Fig. 8(a). (c) The deviation of the data points from the smoothed curve, which corresponds to the sensitivity of this measurement.

Fig. 9. (a) Temperature vs $\dfrac{P_{AC}^{RbFe(MoO_4)_2}}{2\omega\cdot\left|T_{AC}^{RbFe(MoO_4)_2}\right|} - \dfrac{P_{AC}^{Si}}{2\omega\cdot\left|T_{AC}^{Si}\right|}$ in sample 1 of $RbFe(MoO_4)_2$ single crystal. The frequency of temperature oscillation ($2f$) is 400 Hz. The black, blue and red curves are

taken with $P_{AC}$ = 5.8, 8.3 and 16 μW of AC power. The left hand axis depicts the reported specific heat measured by the traditional relaxation methods (green circles).[52] The inset shows $|T_{AC}| \cdot f$ vs $2f$ plot in the sample 1 (black) and 2 (red). (b) $C_p^{RbFe(MoO_4)_2}$ in arbitrary unit up to 20 T (sample 1). Using the temperature oscillation data taken in down-sweeps, these $C_p^{RbFe(MoO_4)_2}$ data were calculated by $\dfrac{P_{AC}^{RbFe(MoO_4)_2}}{2\omega \cdot \left| T_{AC}^{RbFe(MoO_4)_2} \right|} - \dfrac{P_{AC}^{Si}}{2\omega \cdot \left| T_{AC}^{Si} \right|}$ with following Eq. 11. The red, blue, and black curves were measured under 10, 15, and 20 T pulsed magnetic fields, respectively.

Fig.10. (a) Temperature vs magnetic field during the 250 ms pulse in a $Sr_3Cr_2O_8$ single crystal. The frequency of temperature oscillation ($2f$) is 1 kHz. The black, blue and red curves are data of down-sweep, up-sweep, and its smoothed data. (b) Temperature oscillation vs magnetic field. This temperature oscillation data were detected after filtering and amplifying the raw resistance oscillation signal with preamplifier (see the electronic circuit diagram in Fig. 3(b)). The inset is the enlarged figure in the field range of spin ordering transition of $Sr_3Cr_2O_8$. (c) The black curve is $C_p^{Sr_3Cr_2O_8}$ in the arbitrary unit, which corresponding to $\dfrac{P_{AC}^{Sr_3Cr_2O_8}}{2\omega \cdot \left| T_{AC}^{Sr_3Cr_2O_8} \right|} - \dfrac{P_{AC}^{Si}}{2\omega \cdot \left| T_{AC}^{Si} \right|}$ in Eq.11. The $C_p^{Sr_3Cr_2O_8}$ data were calculated from the temperature oscillation data taken in down-sweeps. The open circles are the heat capacity data measured in the DC field[26] and the arrows indicate the dips observed at the field showing MCE. (d) $C_p^{Sr_3Cr_2O_8}$ between 2.3 to 5.6 K. The arrows indicate the peaks in AC-$C_p$.

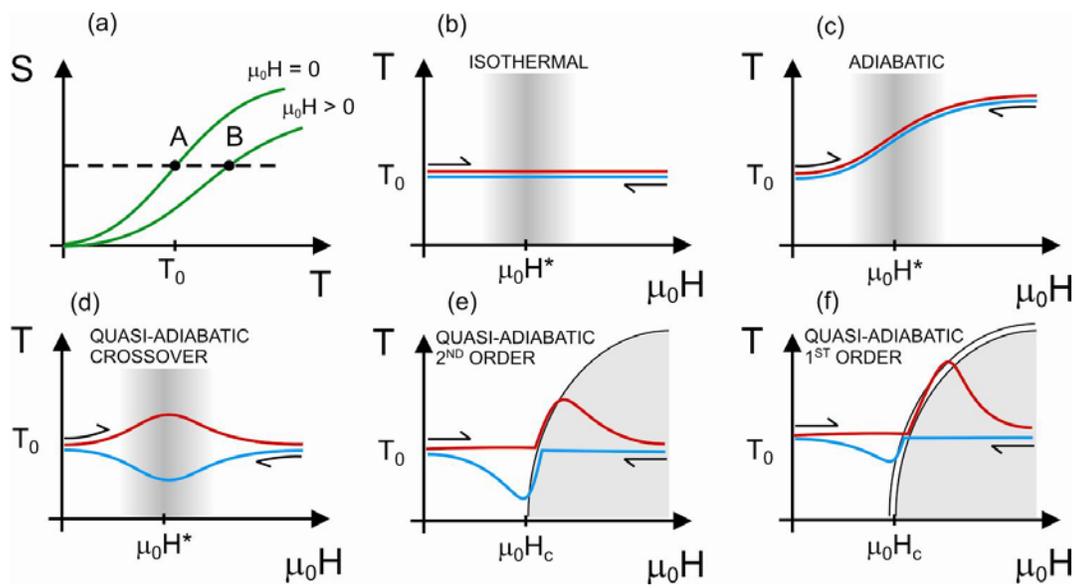

Figure 1

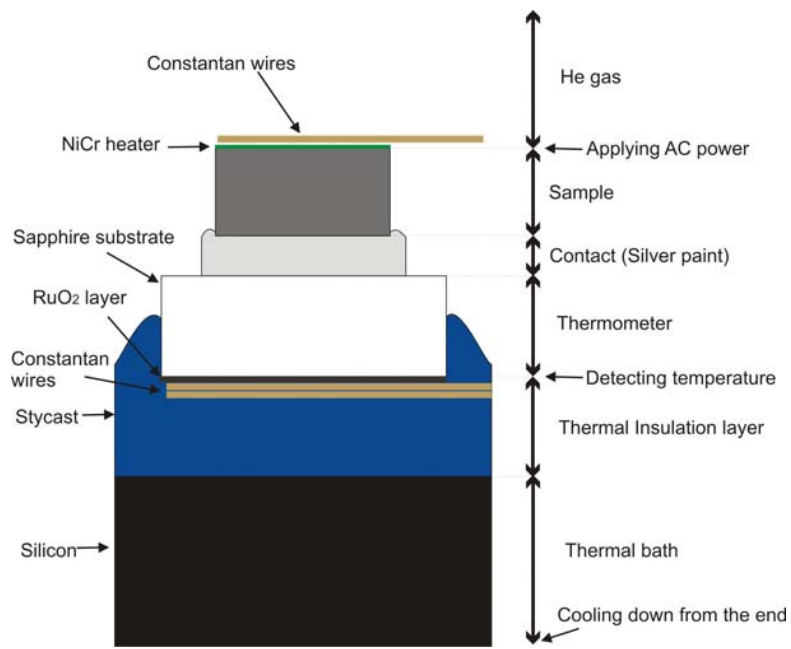

Figure 2

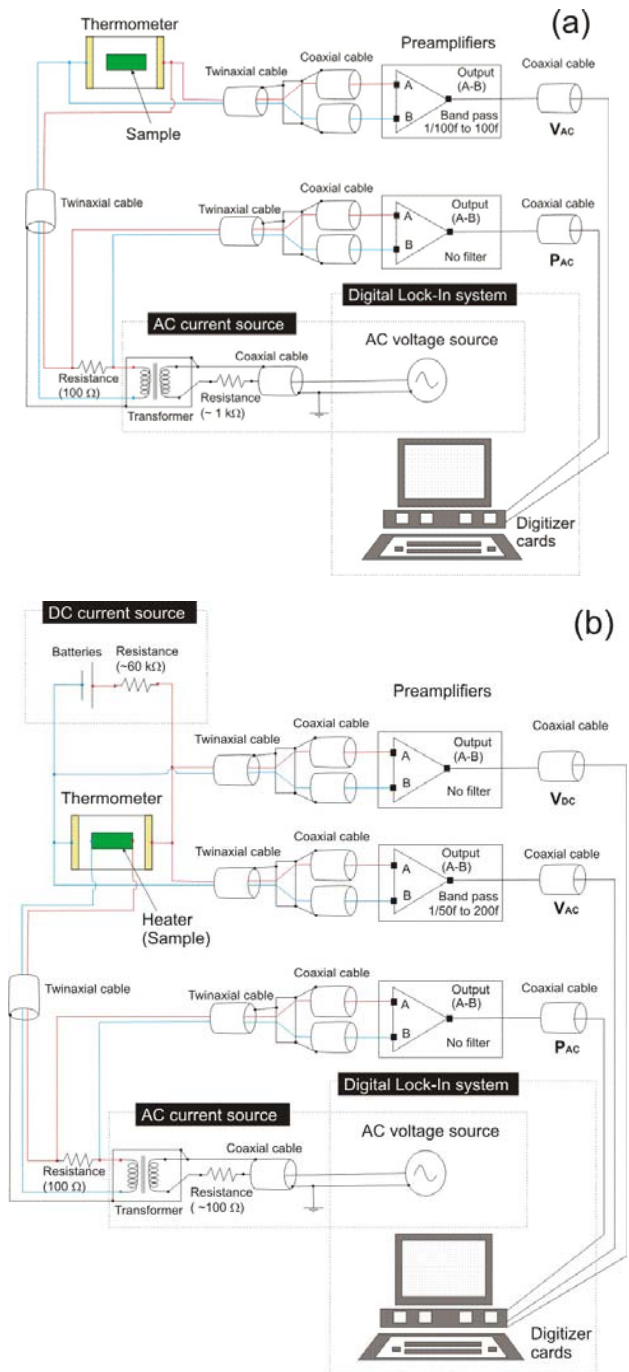

Figure 3

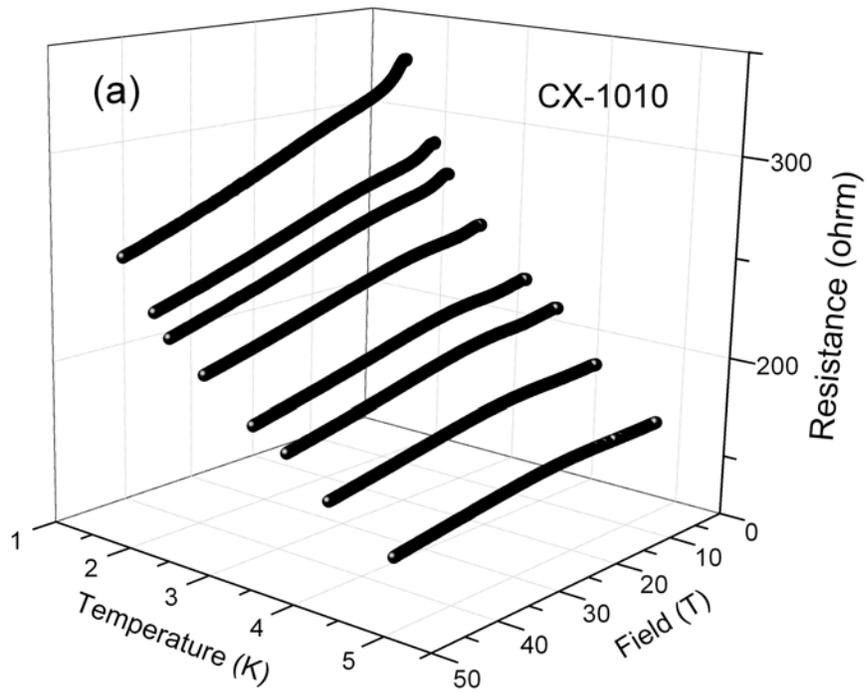

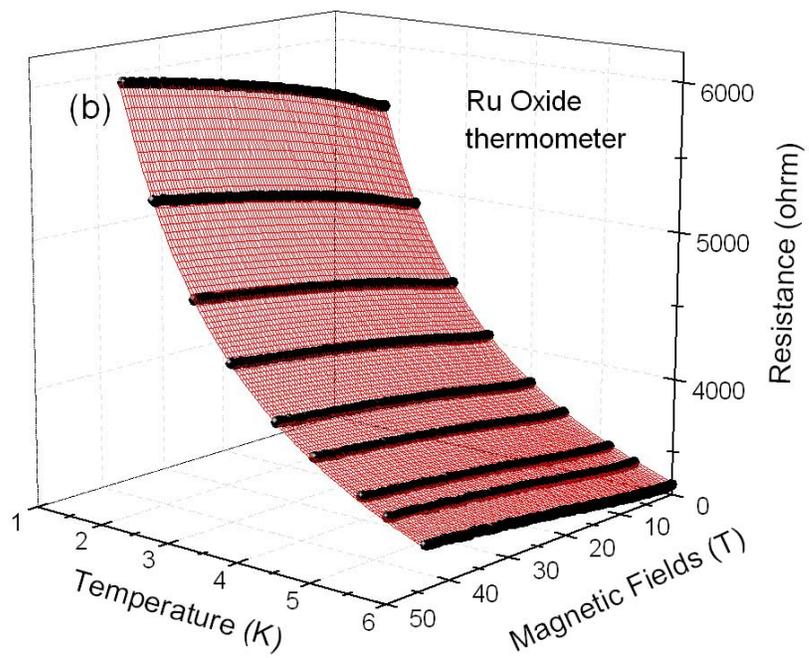

Figure 4

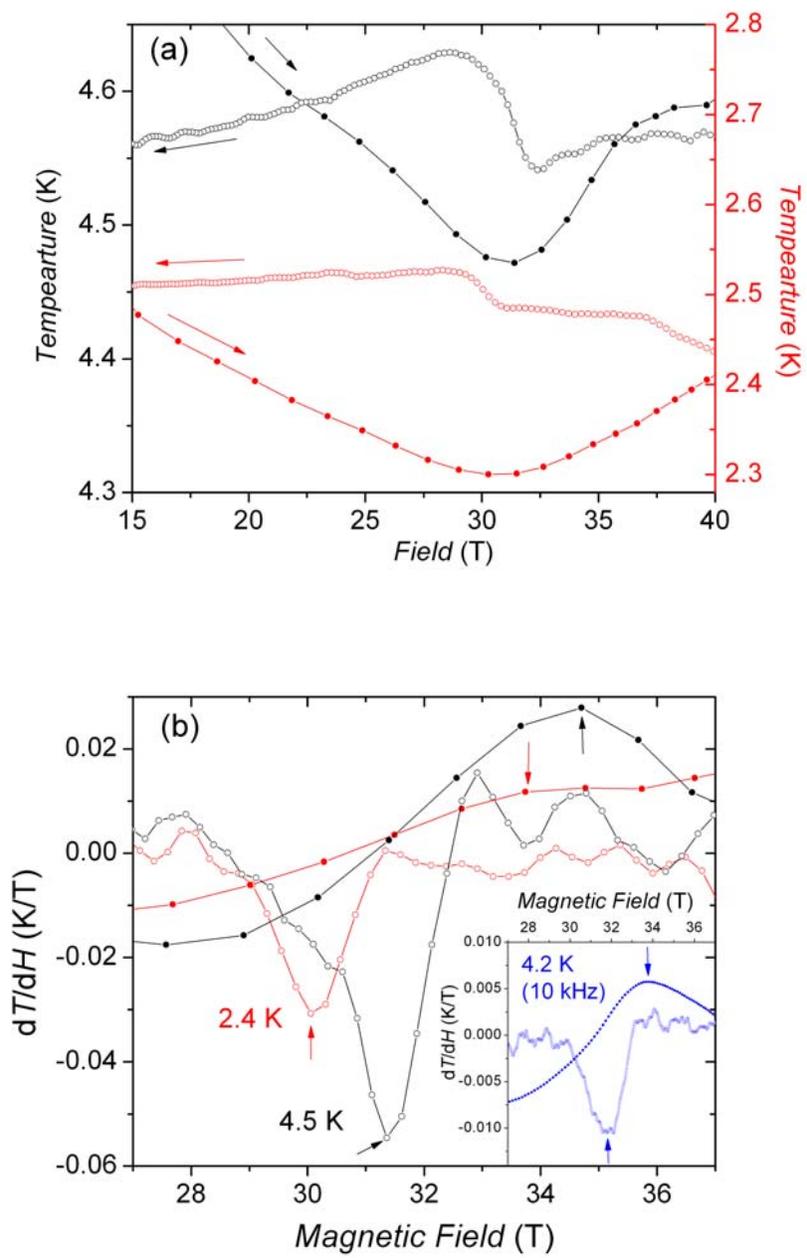

Figure 5

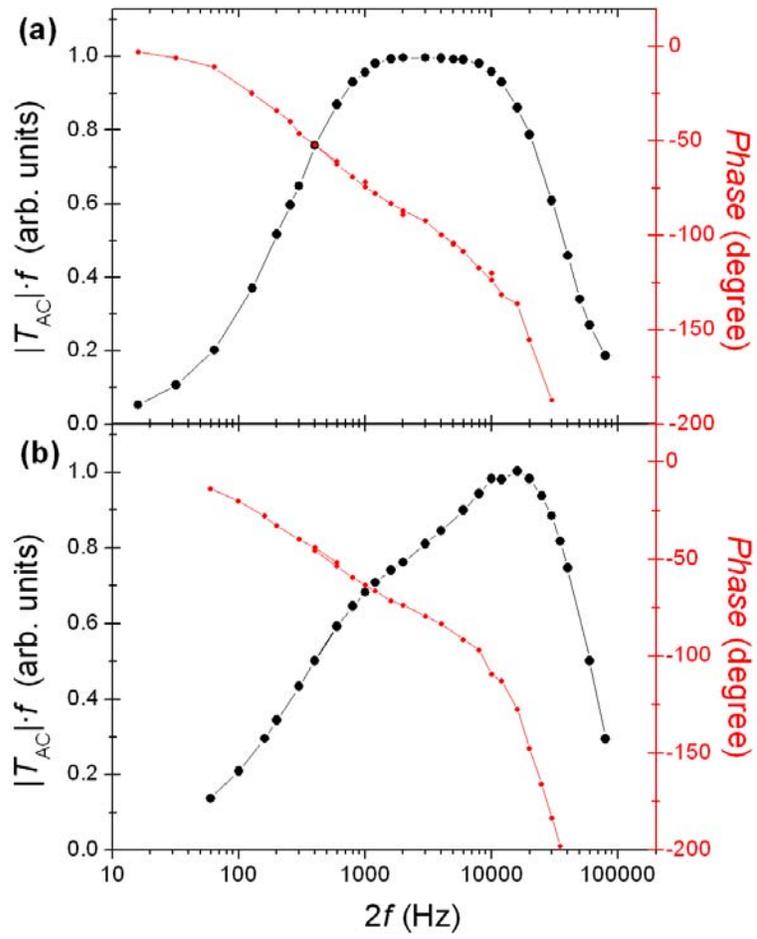

Figure 6

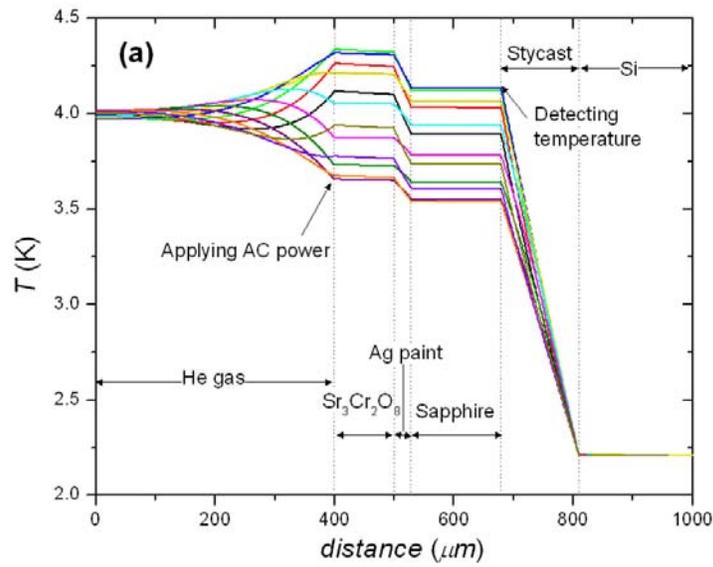

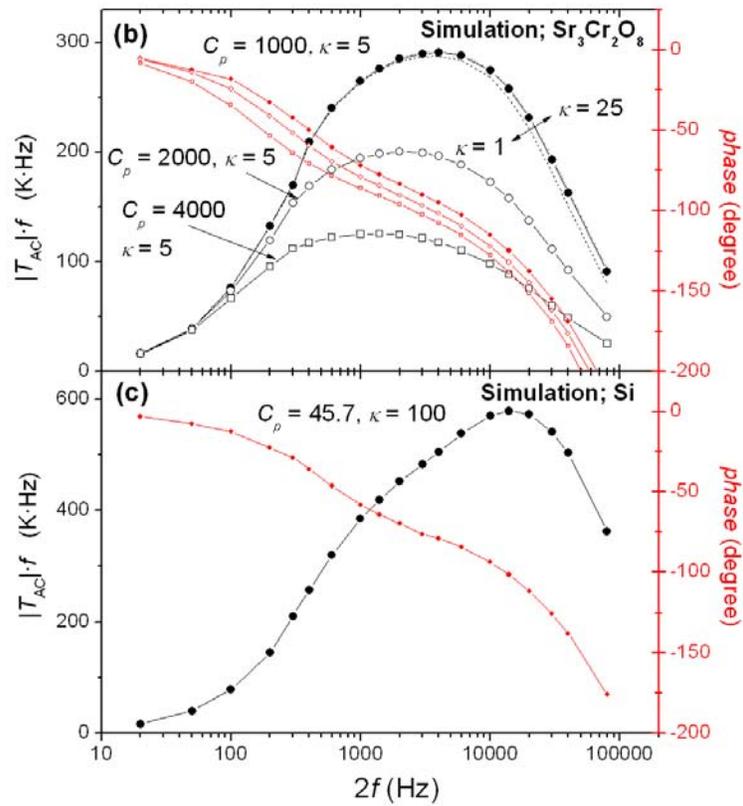

Figure 7

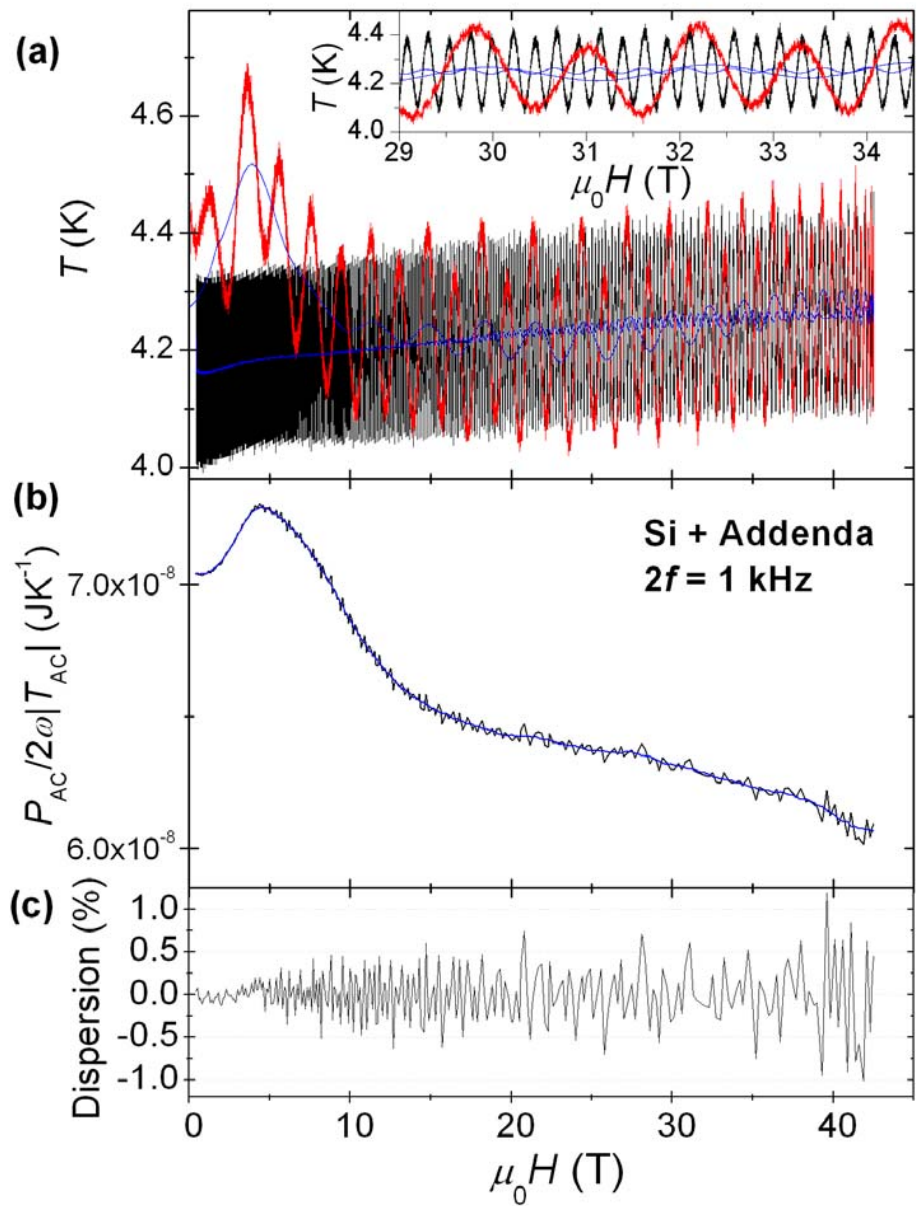

Figure 8

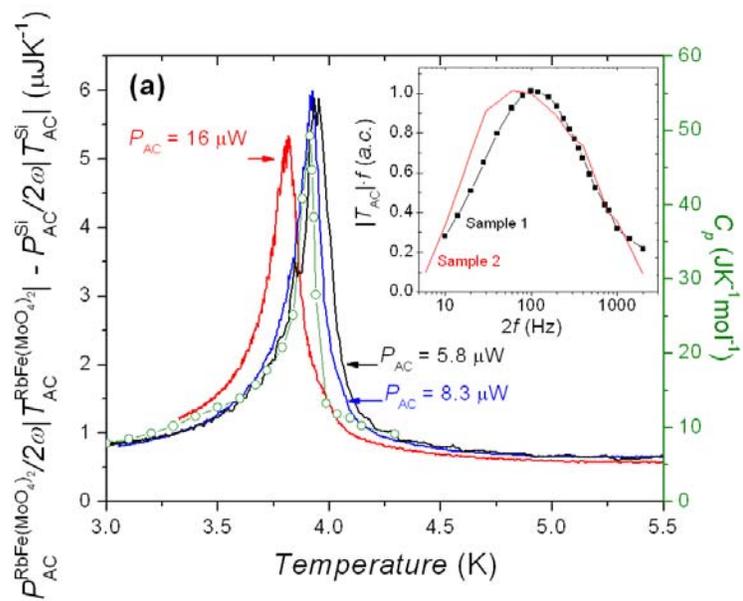

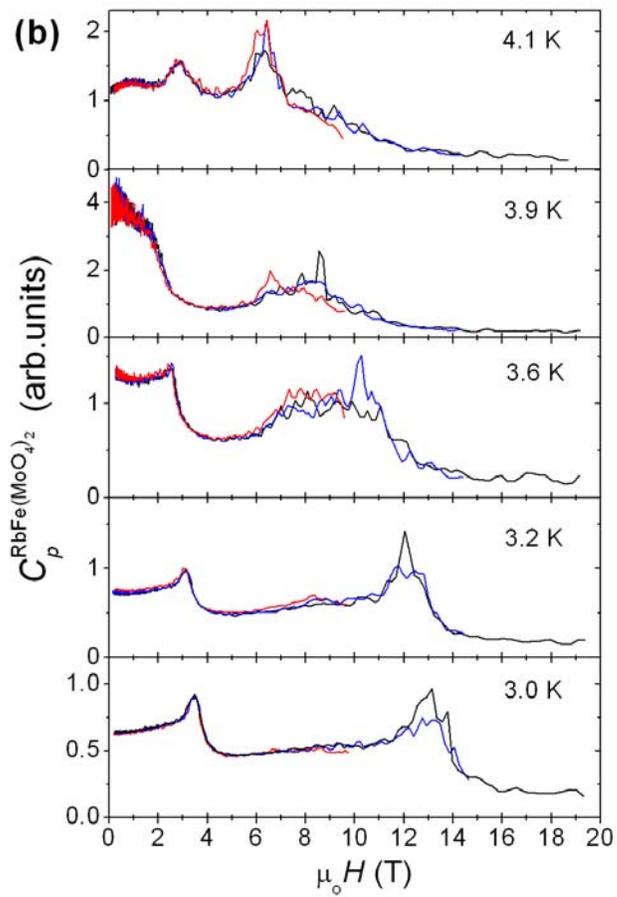

Figure 9

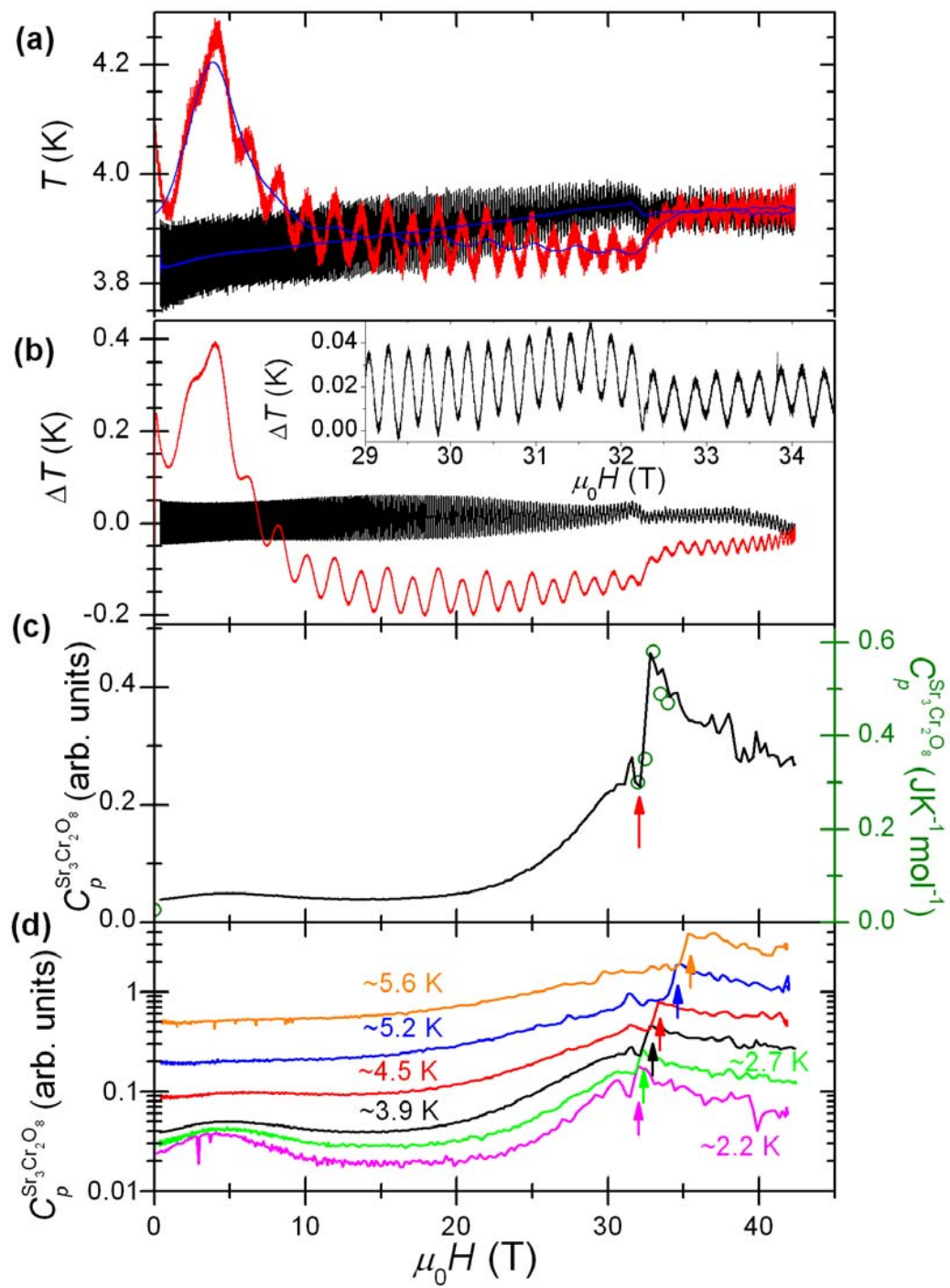

Figure 10